\input phyzzx
\singlespace                     
\nopagenumbers
\headline={\ifodd \pageno\rightheadline \else\leftheadline\fi}
\def\rightheadline{\iffrontpage \nopagenumbers \else \hfill\folio\fi}
\def\leftheadline{\folio\hfill}
\frontpagetrue
\voffset=0.1cm    
\def\boxit#1{\vcenter{\hrule\hbox{\vrule\kern8pt
      \vbox{\kern8pt#1\kern8pt}\kern8pt\vrule}\hrule}}
\def\Boxed#1{\boxit{\hbox{$\displaystyle{#1}$}}} 
\def\bg{\hbox{\fourteeni g}}
\def\sbg{\hbox{\twelvei g}}     
\def\sbeta{\hbox{$\beta \textfont1=\seveni$}}
\def\ssbeta{\hbox{$\beta \textfont1=\fivei$}}
\def\bdelta{\hbox{$\delta \textfont1=\fourteeni$}}
\def\sbdelta{\hbox{$\delta \textfont1=\twelvei$}} 
\def\cg{\hbox{$\cal G\textfont2=\fourteensy$}}
\def\scg{\hbox{$\cal G\textfont2=\twelvesy$}}  
\def\cs{\hbox{$\cal S\textfont2=\fourteensy$}}
\def\scs{\hbox{$\cal S\textfont2=\twelvesy$}}   
\def\ct{\hbox{$\cal T\textfont2=\fourteensy$}}
\def\sct{\hbox{$\cal T\textfont2=\twelvesy$}}   
\def\TableOfContentEntry#1#2#3{\relax}
\def\section#1{\par \ifnum\the\lastpenalty=30000\else
   \penalty-200\vskip\sectionskip \spacecheck\sectionminspace\fi
   \global\advance\sectionnumber by 1
   \xdef\sectionlabel{\the\sectionstyle\the\sectionnumber}
   \wlog{\string\section\space \sectionlabel}
   \TableOfContentEntry s\sectionlabel{#1}
   \noindent {\caps\enspace\bf \sectionlabel\quad #1}\par
   \nobreak\vskip\headskip \penalty 30000
   \ifnum\equanumber<0 \else \global\equanumber=0\fi }
\def\eqname#1{\relax \ifnum\equanumber<0
     \xdef#1{{\noexpand\rm(\number-\equanumber)}}%
       \global\advance\equanumber by -1
    \else \global\advance\equanumber by 1
      \xdef#1{{\noexpand\rm(\sectionlabel.\number\equanumber)}} \fi #1}
\def\AM{Ann. Math.}
\def\CQG{Class. Quantum Grav.}
\def\GRG{Gen. Rel. Grav.}
\def\JMP{J. Math. Phys.}
\def\JSIRAN{J. Sci. I. R. Iran}
\def\NC{Nuovo Cim.}
\def\NP{Nucl. Phys.}
\def\PL{Phys. Lett.}
\def\PR{Phys. Rev.}
\def\PRp{Phys. Rep.}
\def\RMP{Rev. Mod. Phys.}
\REF\farb{Farhoudi, M ``Lovelock tensor as generalized Einstein
          tensor'', {\it gr-qc/9510060}.}
\REF\lovedcbrig{Lovelock, D ``The Einstein tensor and its
                generalizations''
                \journal\JMP&12 (71) 498-501; ``The four dimensionality of
                space and the Einstein tensor'' \journal\JMP&13 (72)
                874-876;\nextline
                Briggs, C C ``Some possible features of general expressions
                for Lovelock tensors and for the coefficients of Lovelock
                Lagrangians up to the $15^{th}$ order in curvature (and
                beyond)'', {\it gr-qc/9808050}.}
\REF\demapoibdehab{Deruelle, N \& Madore, J ``Kaluza--Klein
                   cosmology with
                   the Lovelock Lagrangian'' in {\it Origin and Early History
                   of the Universe}, Proc.~26$^{th}$ Li\`ege
                   Int. Astrophysical Colloquium, July 1986,
                   Belgium, (Cointe-Ougree, Belgium, 1987), pp
                   277-283;\nextline
                   Poisson, E ``Quadratic gravity and the black hole
                   singularity'' \journal\PR&D43 (91) 3923-3928;\nextline
                   Dehghani, M H ``Magnetic branes in Gauss--Bonnet
                   gravity'' \journal\PR&D69 (2004) 064024; ``Accelerated
                   expansion of the universe in Gauss--Bonnet gravity'',
                   {\it hep-th/0404118}.}
\REF\zwizum{Zwiebach, B ``Curvature squared terms and string
            theories''
            \journal\PL&B156 (85) 315-317;\nextline
            Zumino, B ``Gravity theories in more than four dimensions''
            \journal\PRp&137 (86) 109-114.}
\REF\cdttccct{Carroll, S M, Duvvuri, V, Trodden, M \& Turner, M S
              ``Is cosmic speed-up due to new gravitational
              physics?'' \journal\PR&D70 (2004) 043528;\nextline
              Capozziello, S, Cardone, V F, Carloni, S \& Troisi, A ``Curvature
              quintessence matched with observational data'' \journal Int. J. Mod.
              Phys.&D12 (2003) 1969-1982.}
\REF\noodb{Nojiri, S \& Odintsov, S D ``Modified gravity with
           ${\ln} R$ terms and cosmic acceleration'' \journal\GRG&36 (2004)
           1765-1780.}
\REF\dokasowochi{Dolgov, A D \& Kawasaki, M ``Can modified gravity
                 explain accelerated cosmic expansion?'' \journal\PL&B573 (2003)
                 1-4;\nextline
                 Soussa, M E \& Woodard, R P ``The force of
                 gravity from a Lagrangian containing inverse
                 powers of the Ricci scalar'' \journal\GRG&36 (2004) 855-862;\nextline
                 Chiba, T ``$1\over R$ gravity and scalar--tensor
                 gravity'' \journal\PL&B575 (2003) 1-3.}
\REF\nooda{Nojiri, S \& Odintsov, S D ``Modified gravity with
           negative and positive powers of the curvature: unification of the
           inflation and of the cosmic acceleration'' \journal\PR&D68 (2003)
           123512.}
\REF\noodc{Nojiri, S \& Odintsov, S D ``The minimal curvature of
           the universe in modified gravity and conformal anomaly resolution
           of the instabilities'' \journal Mod. Phys. Lett.&A19 (2004)
           627-638.}
\REF\dufa{Duff, M J ``Observations on conformal anomalies''
          \journal\NP&B125 (77) 334-348.}
\REF\fard{Farhoudi, M {\it Non-linear Lagrangian Theories of
          Gravitation}, (Ph.D. Thesis, Queen Mary \& Westfield College,
          University of London, 1995).}
\REF\gra{Graves, J C, {\it The Conceptual Foundation
         of Contemporary
         Relativity Theory}, (MIT Press, 1971), pp 301-302.}
\REF\abbasi{Abbassi, A M {\it Revisiting the Mach's and
            Correspondence Principle in General Relativity and Concept of
            Inertia}, (Ph.D. Thesis, Tarbiat Modarres University, Tehran,
            Iran, 2001).}
\REF\dufc{Duff, M J ``Twenty years of the Weyl anomaly'' \journal\CQG&11
          (94) 1387-1403.}
\REF\dufb{Duff, M J ``Supergravity, Kaluza--Klein and superstrings'' in
          Proc. 11$^{th}$ {\it General Relativity and Gravitation},
          Stockholm, 1986, Ed. M A H MacCallum (Cambridge University
          Press, 1987), pp 18-60.}
\REF\sal{Salam, A ``Gauge unification of fundamental forces''
         \journal\RMP&52 (80) 525-538.}
\REF\mffabsirz{Magnano, G, Ferraris, M \& Francaviglia, M ``Non-linear
               gravitational Lagrangians'' \journal\GRG&19 (87) 465-479;
               ``Legendre transformation and dynamical structure of higher
               derivative gravity'' \journal\CQG&7 (90) 557-570;\nextline
               Sirousse Zia, H ``Singularity theorems and the [general
               relativity + additional matter fields] formulation of metric
               theories of gravitation'' \journal\GRG&26 (94) 587-597.}
\REF\hrpmashabb{Haisch, B, Rueda, A \& Puthoff, H E ``Inertia as a
                zero-point-field Lorentz force'' \journal\PR&A49 (94)
                678-694;\nextline
                Mashhoon, B ``On the origin of inertial accelerations''
                \journal\NC&B109 (94) 187-199;\nextline
                Abbassi, A H \& Abbassi, A M ``A modified theory of
                Newtonian mechanics'' \journal\JSIRAN&7 (96) 277-279.}
\REF\lanb{Lanczos, C ``A remarkable property of the Riemann-Christoffel
          tensor in four dimensions'' \journal\AM&39 (38) 842-850.}
\REF\kono{Kobayashi, S \& Nomizu, K, {\it Foundations of
          Differential
          Geometry}, Vol.~II, (Wiley Interscience, New York, 1969).}
\REF\bida{Birrell, N D \& Davies, P C W, {\it Quantum Fields in
          Curved
          Space}, (Cambridge University Press, 1982).}
\REF\dufaa{Duff, M J ``Inconsistency of quantum field theory in curved
           space-time'' in {\it Quantum Gravity $\Roman 2$: A Second
           Oxford Symposium}, Eds. C J Isham, R Penrose \& D W
           Sciama (Clarendon Press, Oxford, 1981), pp 81-105.}
\REF\bos{Buchbinder, I L, Odintsov, S D \& Shapiro, I L, {\it
         Effective Action in Quantum Gravity}, (Institute of Physics
         Publishing, Bristol and Philadelphia, 1992).}
\REF\stela{Stelle, K S ``Renormalization of higher derivative quantum
           gravity'' \journal\PR&D16 (77) 953-969.}
\REF\chswgswltp{Candelas, P, Horowitz, G T, Strominger, A \& Witten, E
                ``Vacuum configurations for superstrings'' \journal\NP&B258
                (85) 46-74;\nextline
                Green, M B, Schwarz, J H \& Witten, E, {\it Superstring
                Theory}, Vols.~1 \&~2, (Cambridge University Press, 1987);
                \nextline
                Lust, D \& Theusen, S, {\it Lectures on String Theory},
                (Springer, Berlin, 1989);\nextline
                Polchinski, J, {\it String Theory}, (Cambridge University
                Press, 1998).}
\REF\chammuh{Chamseddine, A H ``Topological gauge theory of
             gravity in five
             and all odd dimensions'' \journal\PL&B233 (89)
             291-294;\nextline
             M\"uller-Hoissen, F ``From Chern--Simons to Gauss--Bonnet''
             \journal\NP&B346 (90) 235-252.}
\REF\ddi{Deser, S, Duff, M J \& Isham, C J ``Non-local conformal
         anomalies'' \journal\NP&B111 (76) 45-55.}
\REF\chr{Christensen, S M ``Regularization, renormalization, and covariant
         geodesic point separation'' \journal\PR&D17 (78) 946-963.}
\REF\duf{Duff, M J ``Covariant quantization [of gravity]'' in {\it Quantum
         Gravity: An Oxford Symposium}, Eds. C J Isham, R Penrose
         \& D W Sciama (Clarendon Press, Oxford, 1975), pp 78-135.}
\REF\bcr{Bonora, L, Cotta-Ramusino, P \& Reina, C ``Conformal anomaly
         and cohomology'' \journal\PL&126B (83) 305-308.}
\REF\bpb{Bonora, L, Pasti, P \& Bregola, M ``Weyl cocycles''
         \journal\CQG&3 (86) 635-649.}
\REF\noodog{Nojiri, S, Odintsov, S D \& Ogushi, S
            ``Holographic renormalization group and conformal
            anomaly for AdS$_9$/CFT$_8$\ correspondence''
            \journal\PL&B500 (2001) 199-208.}
\REF\avra{Avramidi, I G ``New algebraic methods for calculating the
          heat
          kernel and effective action in quantum gravity and gauge
          theories'' in {\it Heat Kernel Techniques and Quantum
          Gravity}, Discourses in Mathematics and its Applications, Ed.
          S A Fulling, (Department of Mathematics,
          Texas A\&M University, 1995), pp 115-140, {\it gr-qc/9408028}.}
\REF\farf{Farhoudi, M ``New derivation of Weyl invariants in six
          dimensions'', work in progress.}
\REF\desch{Deser, S \& Schwimmer, A ``Geometric classification of
           conformal anomalies in arbitrary dimensions'' \journal\PL&B309
           (93) 279-284.}
\REF\gilababocsch{Gilkey, P B ``The spectral geometry of a Riemannian
                  manifold'' \journal J. Diff. Geom.&10 (75) 601-618, and
                  ``Recursion relations and the asymptotic behavior of
                  the eigenvalues of the Laplacian'' \journal Compositio
                  Math.&38 (79) 201-240;\nextline
                  Amsterdamski, P, Berkin, A L \& O'Connor, D J ``$b_8$
                  `Hamidew' coefficient for a scalar field'' \journal\CQG&6
                  (89) 1981-1991;\nextline
                  Schimming, R ``Calculation of the heat kernel
                  coefficients'' in {\it Analysis, Geometry and Groups: A
                  Riemann Legacy Volume}, Eds. H M Srivastava \& T M
                  Rassias (Hadronic Press, Palm Harbor, Florida, 1993),
                  pp 627-656.}
\bigskip
\titlestyle{\bf Classical Trace Anomaly}
\bigskip
\centerline{\twelvepoint M Farhoudi\foot{This work was partially
                                         supported by a grant
                                         from the MSRT/Iran.}}
\medskip
\centerline{\it Physics Department,}
\centerline{\it Shahid Beheshti University,}
\centerline{\it Evin, Tehran 1983963113, Iran,}
\centerline{E-mail address: m-farhoudi@sbu.ac.ir}
\bigskip
\noindent PACS number: 04.20-s                     \nextline
Keywords: Higher order gravity; Lovelock \& Einstein tensors; Weyl
          anomaly.
\bigskip
\medskip
\centerline{\bf Abstract}
{\narrower\smallskip \noindent\tenrm\singlespace
We seek an
analogy of the mathematical form of the alternative {\tenit form}
of Einstein's field equations for Lovelock's field equations. We
find that the price for this analogy is to accept the existence of
the trace anomaly of the energy-momentum tensor even in classical
treatments. As an example, we take this analogy to any generic
second order Lagrangian and exactly derive the trace anomaly
relation suggested by Duff. This indicates that an intrinsic
reason for the existence of such a relation should perhaps be,
classically, somehow related to the covariance of the form of
Einstein's equations.
\smallskip}
\section{\bf Introduction}
\indent Recently there has been interest in considering gravity in
higher dimensional space-time. In this context, one may use a
consistent theory of gravity with a more general action, e.g. the
Einstein--Hilbert action plus higher order terms\rlap.\foot{See e.g.
Ref.~[\farb]\ for a brief review of the history of inclusion of
scalar Lagrangians quadratic in the curvature tensor.}\ Especially,
much interest has been in the Lovelock
gravity\rlap,$^{^{[\lovedcbrig]}}$ as the most general second order
Lagrangian which still yields the field equations as second order
equations, and its applications in cosmology, see e.g.
Refs.~[\demapoibdehab] and references therein. This particular
combination of higher order terms are
ghosts--free\rlap.$^{^{[\zwizum]}}$\ Nevertheless, in order to
obtain the observed accelerated cosmic expansion at the present
epoch, other higher order modifications of gravity have also
recently been attracted\rlap,$^{^{[\cdttccct ,\noodb]}}$\ among
which a particular case of $1\over R$ term modification has been
shown to lead to instabilities\rlap.$^{^{[\dokasowochi]}}$\ Though,
Ref.~[\nooda] claims that further modification of this modified
gravity by $R^2$ or other higher terms may resolve the
instabilities, or perhaps make them avoidable\rlap.$^{^{[\noodb
,\noodc]}}$\
\par
In our previous work\rlap,$^{^{[\farb]}}$\ we showed that the
analogy of the Einstein tensor (i.e., the splitting feature of it
into two parts, the Ricci tensor and the term proportional to the
curvature scalar, {\it with} the trace relation between them,
which is also a common feature of {\it each} homogeneous term in
the Lovelock tensor\foot{That is, $G^{(n)}_{\alpha\ssbeta}$ where
the Lovelock tensor, as dimensionally reduction Euler--Lagrange
terms, is$^{^{[\lovedcbrig]}}$
$$
\scg_{\alpha\ssbeta}=-\sum_{0<n<{D\over 2}}\, {1\over 2^{n+1}}\, c_n\,
 \sbg_{\alpha\mu}\,
 \sbdelta^{\mu\alpha_1\ldots\alpha_{2n}}_{\ssbeta\ssbeta_1\ldots\ssbeta_{2n}}
 \, R_{\alpha_1\alpha_2}{}^{\ssbeta_1\ssbeta_2}\cdots
 R_{\alpha_{2n-1}\,\alpha_{2n}}{}^{\ssbeta_{2n-1}\,\ssbeta_{2n}}\equiv
 \sum_{0<n<{D\over 2}}\, c_n\, G^{(n)}_{\alpha\ssbeta}\ ,\eqn\lovet$$
and where the cosmological term has been neglected,
$G^{(1)}_{\alpha\ssbeta}=G_{\alpha\ssbeta}$ the Einstein tensor,
$\sbdelta^{\alpha_1\ldots\alpha_p}_{\ssbeta_1\ldots\ssbeta_p}$ is
the generalized Kronecker delta symbol, which is identically zero
if $p>D$, and the maximum value of $n$ is related to the dimension
of space-time by
$$
n_{_{\rm max.}}\!=\cases{{D\over 2}-1&even $D$\cr
         {{D-1}\over 2}&odd $D$\ .\cr}\eqn\nlim$$
Also, our conventions are a metric of signature $+2$,\
$R^{\mu}{}_{\nu\alpha\ssbeta}=-\Gamma^{\mu}{}_{\nu\alpha,\,\ssbeta}+\cdots$,
$R_{\mu\nu}\equiv R^{\alpha}{}_{\mu\alpha\nu}$, and the
homogeneity is taken with respect to the metric and its
derivatives with the homogeneity degree number (HDN) conventions
of ${}^{[+1]}\sbg^{\mu\nu}$ and
${}^{[+1]}\sbg^{\mu\nu}{}_{,\alpha}$ (see Ref.~[\farb] for
details). Hence, one can relate the orders $n$ in any Lagrangian,
as in $L^{(n)}$, that represents its HDN.})
can be generalized, via a generalized trace operator (which we defined
and denoted by {\sl Trace} for homogeneous tensors\foot{That is, for a
 general $\bigl({N\atop M}\bigr)$ tensor which is a homogeneous function of
 degree $h$ with respect to the metric and its derivatives, we defined
 $$
 {\rm Trace}\, {}^{[h]}A^{\alpha_1\ldots\alpha_N}
    {}_{\ssbeta_1\ldots\ssbeta_M}
    :=\cases{{1\over h-{N\over 2}+{M\over 2}}\, {\rm trace}\,
    {}^{[h]}A^{\alpha_1\ldots\alpha_N}{}_{\ssbeta_1\ldots\ssbeta_M}&\quad
    when $h-{N\over 2}+{M\over 2}\not=0$\cr
    {\rm trace}\, {}^{[h]}A^{\alpha_1\ldots\alpha_N}
    {}_{\ssbeta_1\ldots\ssbeta_M}&\quad when
    $h-{N\over 2}+{M\over 2}=0$\ .\cr}\eqn\Tracedu$$
 Hence, for example, when $h\neq 0$ and $h'\not=0$, one gets
 $$
 {\rm Trace}\,\bigl({}^{[h']}C\, {}^{[h]}A_{\mu\nu}\bigr)=
    \cases{{h+1\over h'+h+1}\, {}^{[h']}C\,{\rm Trace}\,
           {}^{[h]}A_{\mu\nu}&\quad for $h\not=-1$\cr
           {1\over h'}\, {}^{[h']}C\,{\rm Trace}\,
           {}^{[h]}A_{\mu\nu}&\quad for $h=-1$\ .\cr}\eqn\Tracexampl$$}),
to any inhomogeneous Euler--Lagrange expression if it can be spanned
linearly in terms of homogeneous tensors.
\par
As an example, we demonstrated this analogy of the mathematical
form of the Einstein tensor for the Lovelock tensor, and showed
that it can be written as
$$
\cg_{\alpha\sbeta}=\Re_{\alpha\sbeta}-{1\over 2}\bg_{\alpha\sbeta}\,\Re
                                                          \ ,\eqn\elove$$
with the relation
$$
{\rm Trace}\, \Re_{\alpha\sbeta}=\Re\ ,\eqn\grt$$
where
$$
\Re_{\alpha\sbeta}\equiv \sum_{0<n<{D\over 2}}\, c_n\, R^{(n)}_{\alpha\sbeta}
   \qquad ,\qquad
   \Re=\kappa^2\, {\cal L}\equiv \sum_{0<n<{D\over 2}}\, c_n\,
   R^{(n)}=\kappa^2 \sum_{0<n<{D\over 2}}\, c_n\, L^{(n)}
   \ ,\eqn\rrlove$$
where ${\cal L}$ is the Lovelock Lagrangian, and
$R^{(n)}_{\alpha\sbeta}$ and $R^{(n)}$ are defined as
$$
R^{(n)}_{\alpha\sbeta}\equiv {n\over 2^n}\,\bdelta^{\alpha_1\alpha_2%
   \ldots\alpha_{2n}}_{\alpha\ \sbeta_2\ldots\sbeta_{2n}}\,
   R_{\alpha_1\alpha_2\sbeta}{}^{\sbeta_2}\, R_{\alpha_3\alpha_4}
   {}^{\sbeta_3\sbeta_4}\cdots R_{\alpha_{2n-1}\,\alpha_{2n}}
   {}^{\sbeta_{2n-1}\,\sbeta_{2n}}\eqn\ricn$$
and
$$
R^{(n)}\equiv {1\over 2^n}\,
  \bdelta^{\alpha_1\ldots\alpha_{2n}}_{\sbeta_1\ldots\sbeta_{2n}}\,
  R_{\alpha_1\alpha_2}{}^{\sbeta_1\sbeta_2}\cdots
  R_{\alpha_{2n-1}\,\alpha_{2n}}{}^{\sbeta_{2n-1}\,\sbeta_{2n}}
  \ ,\eqn\rn$$
where also $R^{(1)}_{\alpha\sbeta}\equiv R_{\alpha\sbeta}$,\
$R^{(1)}\equiv R$,\
$\kappa^2\equiv {16\pi G\over c^4}$,\
$c_1\equiv 1$ and the other $c_n$ constants are of the
order of Planck's length to the power $2(n-1)$.
\par
Hence, we stated the Lovelock gravitational field equations in the
form of
$$
\cg_{\alpha\sbeta}={1 \over 2}\kappa^2\, T_{\alpha\sbeta}\ ,\eqn\lovee$$
and classified the Lovelock tensor/Lagrangian as a {\it
generalized} Einstein tensor/\break
Lagrangian, and called
$\Re_{\alpha\sbeta}$ and $\Re$ {\it generalized} Ricci tensor and
{\it generalized} curvature scalar, respectively.
\par
Now in this work, still motivated by the principle of general covariance,
we proceed further the analogy and also enforce the mathematical form of the
alternative {\it form} of Einstein's field equations (as a {\it covariant
form} for gravitational field equations) for the relevant alternative form
of Lovelock's field equations. From this, we find that the price for this
analogy is to accept the existence of the trace anomaly of the
energy-momentum tensor even in classical treatments.
Then, as an example, we take this analogy to any generic second order
Lagrangian and exactly derive the trace anomaly relation suggested by
Duff\rlap.$^{^{[\dufa]}}$\
\par
Besides the very well known classical successes of Einstein's theory, the
above analogy is a part of a programme to impose an analogy of Einstein's
gravity on Lovelock's gravity\rlap,$^{^{[\fard]}}$\ wherein the latter is
then considered as a generalized Einstein's gravity. Actually, because of the
non-linearity of the field equations, it is very difficult to find out
non-trivial exact analytical solutions of Einstein's equation with higher
curvature terms. So, our ultimate proposal has been to construct, and hence
to achieve, a generalized counterpart for each essential term used in
Einstein's gravity, especially, the metric. This, we believe, would give a
better view on higher order gravities, and would also let straightway to
apply the results of Einstein's theory to Lovelock's theory.
\par
In Section 2, we will review the alternative form of the Einstein
field equations; and in Section~3, we will perform the same
analogy in order to get the alternative form of the Lovelock field
equations. In Section~4, we will perform some discussions mainly
on an idea that the geometrical curvature inducing matter.
Finally, in Section~5, as an example, we will take this analogy to
any generic second order Lagrangian and derive Duff's
suggested$^{^{[\dufa]}}$ relation for the trace anomaly.
\section{\bf Alternative Form of Einstein's Field Equations}
\indent
With an amount of manipulation one can put Einstein's
field equations, $G_{\alpha\sbeta}={1 \over 2}\kappa^2\,
T_{\alpha\sbeta}$, into an alternative {\it form}. Contracting the
indices (i.e. calculating the standard trace or the {\it Trace},
because the HDN of the Einstein tensor is
zero)\rlap,$^{^{[\farb]}}$\ one gets
$$
{\rm Trace}\, G_{\alpha\sbeta}=\Bigl(1-{D\over 2}\Bigr)R
                              ={1 \over 2}\kappa^2\, T\ .\eqn\tgo$$
Obtaining $R$ from this equation and substituting it back into
Einstein's equations, one gets the equivalent forms of Einstein's field
equations,
$$
R_{\alpha\sbeta}={1\over 2}\kappa^2\Bigl(T_{\alpha\sbeta}
        -{1\over D-2}\bg_{\alpha\sbeta}\, T\Bigr)\equiv
        {1\over 2}\kappa^2\, S_{\alpha\sbeta}\qquad\ {\rm when}\
        D\not=2\ ,\eqn\eea$$
where, for convenience, we refer to $S_{\alpha\sbeta}$ as {\it Source}
tensor. The word {\it form} has been applied for the alternative
gravitational field equations because of the pedagogical reasons, mainly,
in order to emphasize that $R_{\mu\nu}$ is not the gravitation tensor and
$S_{\mu\nu}$ is not the source term since otherwise they do not satisfy the
requirement of vanishing covariant derivatives.
\par
In a vacuum (where there is no matter, i.e. when $T_{\mu\nu}=0$), one
gets $R_{\mu\nu}=0$. But, in four or more dimensions, this does not mean
flat space-time, and gravitational fields exist in empty space. That is, in
the general relativity, the space-time has itself some essence independent
of matter, contrary to the strong version of the Mach
idea\rlap.$^{^{[\gra]}}$
The link to this issue, in our opinion, could be the relation between $T$
and $R$ as in equation \tgo , and the basic concept of matter.
This is somehow a procedure that it may indicate a more compatibility
between the general relativity and the strong version of the Mach
idea\rlap.$^{^{[\abbasi]}}$
\par
From the definition of the Source tensor,
it is evident that vacuum is also equivalent to the case when $S_{\mu\nu}$
vanishes, as equation \eea\ then yields $R_{\mu\nu}=0$.
\section{\bf Alternative Form of Lovelock's Field Equations}
\indent
We will follow the same route of the previous section to get an
alternative form of Lovelock's field equations. However, for this case with
either the trace or the {\it Trace}, one cannot easily substitute for $\Re$
in terms of the trace or the {\it Trace} of $T_{\alpha\sbeta}$, as to be
shown in the following. Therefore, its alternative form will differ
(slightly) from Einstein's one.
\par
Taking the {\it Trace} of \lovee , and using \grt\ and \Tracedu ,
one gets
$$
{\rm Trace}\,\cg_{\alpha\sbeta}=\sum_{0<n<{D\over 2}}
  \Bigl(1-{D\over 2n}\Bigr) c_n\, R^{(n)}={1 \over 2}\kappa^2\,{\rm Trace}\,
  T_{\alpha\sbeta}\ .\eqn\TrG$$
Obviously, one cannot extract $\sum c_n\, R^{(n)}$, as $\Re$, out of the
summation in the above equation.
\par
By setting ${\sl Trace}\, T_{\alpha\sbeta}\equiv T$, from
equation~\TrG\ we have
$$
T=-\kappa^{-2}\,\sum_{0<n<{D\over 2}}\Bigl({D\over n}-2\Bigr)
                 c_n\, R^{(n)}\ .\eqn\TraceG$$
\par
A similar form of equation \eea\ for Lovelock's field equations
can now be found by the following procedure.
\par
By adding and subtracting a term in equation \TrG\ we obtain
$$
\sum_{0<n<{D\over 2}}\Bigl(1-{D\over 2}\Bigr) c_n\, R^{(n)}+
       \sum_{0<n<{D\over 2}}\Bigl({D\over 2}-{D\over 2n}\Bigr) c_n\, R^{(n)}
      ={1 \over 2}\kappa^2\, T\ .$$
Substituting for the first term from equation \rrlove\ and solving
for $\Re$ when $D\not=2$, it yields
$$
\Re=-{\kappa^2\over D-2}\, T+{D\over D-2}\,\sum_{0<n<{D\over 2}}
   \,{n-1\over n}\, c_n\, R^{(n)}\ .\eqn\rgn$$
By putting this into the field equations \lovee , we finally get
$$
\Re_{\alpha\sbeta}
        -{D\over 2\bigl(D-2\bigr)}\,\bg_{\alpha\sbeta}
        \,\sum_{0<n<{D\over 2}}
        \,{n-1\over n}\, c_n\, R^{(n)}={1\over 2}\kappa^2
        \, S_{\alpha\sbeta}\qquad\ {\rm when}\ D\not=2\ .\eqn\geea$$
\par
This is an analogous form of equation \eea\ that can be achieved
in this stage. However, by continuing this procedure below, one
can write equation \geea\ in a manifestly more analogous form with
equation \eea . At present, one should notice that in the second
term of the left hand side of the above equation, there is no
contribution for the case $n=1$, i.e. the associated Einstein's
part\rlap.\foot{Almost the same thing is obtained if one considers
${\sl trace}\,\scg_{\alpha\ssbeta}$ instead. Taking the trace of
equation \lovee , and using ${1\over n}\,{\rm trace}\,
R^{(n)}_{\alpha\sbeta}=R^{(n)}$, one has
$$
{\rm trace}\,\scg_{\alpha\ssbeta}=\sum_{0<n<{D\over 2}}
  \bigl(n-{D\over 2}\bigr) c_n\, R^{(n)}={1 \over 2}\kappa^2\,{\rm trace}\,
  T_{\alpha\ssbeta}\ ,$$
and then, one gets
$$
\Re_{\alpha\ssbeta}-{1\over D-2}\,\sbg_{\alpha\ssbeta}\,
  \sum_{0<n<{D\over 2}}\, (n-1)\, c_n\, R^{(n)}={1\over 2}\kappa^2\bigl(
  T_{\alpha\ssbeta}-{1\over D-2}\,\sbg_{\alpha\ssbeta}\,{\rm trace}\,
  T_{\alpha\ssbeta}\bigr)\ ,$$
where its difference with equation \geea\ is due to the difference between
${\sl trace}\, T_{\alpha\ssbeta}\equiv T^\mu{}_\mu$ and
${\sl Trace}\, T_{\alpha\ssbeta}\equiv T$, when used for dealing with
Lovelock's field equations (i.e., $\scg_{\alpha\ssbeta}$ is an inhomogeneous
tensor).}\
Comparing the appearance of equation \geea\ (or even equation \lovee ) with
Einstein's equation, it is likely (as an analogous demand) that its second
term on the left hand side is responsible for changes in the trace of
energy-momentum tensor.
\par
Our purpose is to eliminate the case of $n=2$ from the second term in the
left hand side of equation \geea\ by the same procedure that we performed
for equation \lovee\ which then led to equation \geea . We do this in the
following manipulation.
\par
Using the following relation
$$
{n-1\over 2n}={(n-1)^2\over n^2}-{(n-1)(n-2)\over
2n^2}\eqn\nrela$$
in equation \geea , we have
$$
\Re_{\alpha\sbeta}-{D\over D-2}\,\bg_{\alpha\sbeta}\,\sum_{0<n<{D\over 2}}
  \biggl[{(n-1)^2\over n^2}-{(n-1)(n-2)\over 2n^2}\biggr]c_n\, R^{(n)}
  ={1\over 2}\kappa^2\, S_{\alpha\sbeta}\ .\eqn\geeb$$
\par
Taking the {\it Trace} of \geea , and using equations \grt\ and
\Tracexampl , we get$^{^{[\farb]}}$
$$
\sum_{0<n<{D\over 2}}\Bigl(1-{D^2\over D-2}\times {n-1\over 2n^2}\Bigr)%
        c_n\, R^{(n)}={1\over 2}\kappa^2\, S\ ,\eqn\strace$$
where $S\equiv {\sl Trace}\, S_{\alpha\sbeta}$\rlap.\foot{Note
that, if one uses this $S$ and the definition of
$S_{\alpha\ssbeta}$, equation \eea , then by the aid of equation
\TraceG , one gets$^{^{[\farb]}}$
$$
\eqalign{S
  &=T-{-\kappa^{-2}\over D-2}\sum_{0<n<{D\over 2}}\Bigl({D\over n}-2\Bigr)
   c_n\, {\rm Trace}\Bigl(\sbg_{\alpha\ssbeta}\, R^{(n)}\Bigr)\cr
  &=-\kappa^{-2}\sum_{0<n<{D\over 2}}\Bigl({D\over n}-2\Bigr)
   \Bigl[1-{D\over n(D-2)}\Bigr]c_n\, R^{(n)}\ ,\cr}\eqn\traceS$$
which is obviously identical to equation \strace .}\
\par
For the first term of equation \strace , we substitute from
equation \rgn , while using relation \nrela , and for its second
term we employ
$$
{n-1\over n^2}={(n-1)^2\over n^2}-{(n-1)(n-2)\over n^2}\
.\eqn\nrelb$$ Then, after rearrangements of equation \strace , we
get
$$
\eqalign{-{D\over D-2}\sum_{0<n<{D\over 2}} {(n-1)^2\over n^2}\,
     c_n\, R^{(n)}=
    &{\kappa^2\over D-4}\, S
     +{2\kappa^2\over \bigl(D-2\bigr)\bigl(D-4\bigr)}\, T\cr
    &-{D\over D-4}\sum_{0<n<{D\over 2}} {(n-1)(n-2)\over n^2}\,
     c_n\, R^{(n)}\ ,\cr}\eqn\rgna$$
when $D\not=4$.
\par
By substituting equation \rgna\ for the second term of equation
\geeb , we finally find that
$$
\eqalign{\Re_{\alpha\sbeta}
   &-{D^2\over 2\bigl(D-2\bigr)\bigl(D-4\bigr)}\,\bg_{\alpha\sbeta}
    \,\sum_{0<n<{D\over 2}}{(n-1)(n-2)\over n^2}\, c_n\, R^{(n)}\cr
   &\ ={1\over 2}\kappa^2\, S_{\alpha\sbeta}
    -{\kappa^2\over D-4}\,\bg_{\alpha\sbeta}\, S
    -{2\kappa^2\over \bigl(D-2\bigr)\bigl(D-4\bigr)}\,\bg_{\alpha\sbeta}\, T
    \equiv {1\over 2}\kappa^2\, S^{(2)}_{\alpha\sbeta}\ ,
    \qquad\ \eqname\geec\cr}$$
where, by using the definition of $S_{\alpha\sbeta}$, equation
\eea , we have
$$
S^{(2)}_{\alpha\sbeta}=T_{\alpha\sbeta}-{1\over D-2}\,
           \bg_{\alpha\sbeta}\, \Bigl(T+T^{(2)}\Bigr)\ ,\eqn\dst$$
where
$$
T^{(2)}\equiv {1\over D-4}\Bigl[2\bigl(D-2\bigr)S+4\, T\Bigr]\
.\eqn\rdtt$$ Or equivalently, by substituting for $S$ and $T$ in
the above equation from equations \traceS\ and \TraceG\
respectively, we obtain
$$
T^{(2)}=-{2\kappa^{-2}\, D\over D-4}
      \sum_{0<n<{D\over 2}}{n-1\over n}
      \Bigl({D\over n}-2\Bigr)c_n\, R^{(n)}\ .\eqn\dtt$$
\par
In this notation, $S^{(2)}_{\alpha\sbeta}$ and $T^{(2)}$ give our desired
result \undertext{up to the $n_{_{\rm max.}}=2$}. Hence, we have also chosen
$S^{(1)}_{\alpha\sbeta}\equiv S_{\alpha\sbeta}$, but $T^{(1)}\equiv 0$.
\par
Equation \geec\ is similar to equation \geea , however, in the
second term of the left hand side there is no contribution for the
cases of $n=1$ and $n=2$. If we continue this iteration to all
orders of $n$ we eventually get
$$
\Re_{\alpha\sbeta}={1\over 2}\kappa^2\, S^{(n_{_{\rm max.}})}_{\alpha\sbeta}
                                                     \ ,\eqn\ngeea$$
where
$$
S^{(n_{_{\rm max.}})}_{\alpha\sbeta}\equiv T_{\alpha\sbeta}
   -{1\over D-2}\bg_{\alpha\sbeta}\, \ct\eqn\dgs$$
with
$$
\ct\equiv T+T^{(n_{_{\rm max.}})}=-\kappa^{-2}\,\bigl(D-2\bigr)\Re
                                                     \ ,\eqn\dgt$$
and
$$
T^{(n_{_{\rm max.}})}=-\kappa^{-2}\,D\,\sum_{0<n<{D\over 2}}
      \,{n-1\over n}\, c_n\, R^{(n)}\ .\eqn\dtn$$
\par
Of course, one can also go from equation \ngeea\ back to equation \lovee ,
so \lovee\ and \ngeea\ should be regarded as entirely equivalent forms
of Lovelock's field equations. Equation \ngeea\ is the closest similar form
of equation \eea\ that we have reached by iteration of the above procedure.
But after all, it is clear that we have considered the second term in
equation \geea\ as {\it equivalent} to the {\sl Trace} of an
energy-momentum tensor which we have tried to justify by this procedure.
Indeed, this is (almost) the same as the case of the alternative form of
Einstein's gravitational theory, and the same argument as was used there to
clarify the word {\it form} is also applied here. Although here, the
definition of each of $S^{(n)}_{\alpha\sbeta}$'s, except
$S^{(n_{_{\rm max.}})}_{\alpha\sbeta}$, is not unique.
\par
We know that $T$ is the {\it Trace} of the energy-momentum tensor, an
entity independent of the gravitational characters, whose relation to the
geometry is through the gravitational field equations as given by
field equation \TraceG .
But, what is $T^{(n_{_{\rm max.}})}$? Does it also have an entity
independent of the
gravitational characters? Or is it a symbol defined by equation \dtn ?
\par
We obtained $T^{(n_{_{\rm max.}})}$ in the same way that we got $T^{(2)}$,
where the latter is defined by equation \rdtt\ with respect to $T$ and $S$,
i.e. things that depend on $T_{\alpha\sbeta}$. So $T^{(n_{_{\rm max.}})}$
should also
have the same kind of relation to $T_{\alpha\sbeta}$, independently of the
gravitational characters. But, we still cannot know its exact relation.
For example, for $T^{(2)}$, one gets into difficulties once one wants
to find out its exact relation with $T_{\alpha\sbeta}$. This is because as,
$$
S={\rm Trace}\, S_{\alpha\sbeta}=T-{1\over D-2}\,{\rm Trace}
                                  \bigl(\bg_{\alpha\sbeta}\, T\bigr)\ ,$$
in order to proceed we need to substitute for $T$ from equation
\TraceG\ (and then use the distributivity of the {\it Trace}).
Otherwise, if one just considers the fact that $T_{\alpha\sbeta}$,
and so $T$, does not depend on the metric (and hence its HDN is
zero), one gets
$$
S={-2\, T\over D-2}={2\kappa^{-2}\over D-2}\sum_{0<n<{D\over 2}}
                \Bigl({D\over n}-2\Bigr)c_n\, R^{(n)}\ ,\eqn\wrongs$$
where we used equation \TraceG\ in the last step.
Whereas, if we use equation \TraceG\ before evaluating the {\it Trace},
we have (from equation \traceS ):
$$
S={2\kappa^{-2}\over D-2}\sum_{0<n<{D\over 2}}\Bigl({D\over n}-2\Bigr)
      c_n\, R^{(n)}-{D\kappa^{-2}\over D-2}\sum_{0<n<{D\over 2}}
      \Bigl({D\over n}-2\Bigr)\Bigl(1-{1\over n}\Bigr)c_n\, R^{(n)}\ .$$
Obviously, only in the case of $n=1$, i.e. Einstein's gravity, is the above
equation equal to equation \wrongs .
\par
Besides, we must warn that in the above procedure, wherever we have used
equation \TraceG\ (or similar ones) for $T$ in order to evaluate a relevant
{\sl Trace} (e.g. ${\sl Trace}\bigl(\bg_{\alpha\sbeta}\, T\bigr)$), this is
not obviously valid when $T$ vanishes. These equations are \strace ,
\traceS , \rgna , \geec , \dtt , \ngeea , \dtn\ and the second part of
\dgt . Therefore, the definitions of \dst , \rdtt , \dgs\ and the first part
of \dgt\ are also undefined when $T_{\alpha\sbeta}$ (hence $T$) vanishes.
That is, using the above procedure, we do not know the real values of
$S^{(2)}_{\alpha\sbeta}$, $T^{(2)}$, $S^{(n_{_{\rm max.}})}_{\alpha\sbeta}$
and $T^{(n_{_{\rm max.}})}$ in a vacuum.
\par
On the other hand, in Einstein's gravitational theory, by equation
\tgo\ we have that $R\propto T$. Hence, as an {\it analogous
demand}, in Lovelock's gravitational theory we should get
$\Re\propto \ct$. Indeed, we have given this proportionality
above, equation \dgt , where we also showed that $\ct$ is not
equal to ${\sl Trace}\, T_{\alpha\sbeta}$, but can be equal to
$T+T^{(n_{_{\rm max.}})}$. But, from now on, we waive the details
of the foregoing procedure, and base our reasoning on the latter
analogous demand. Hence, we take $\ct$ as {\it equivalent} to
$\Re$ with the same ratio of equation \dgt . We also use $T'$
instead of $T^{(n_{_{\rm max.}})}$ (i.e., $T'\equiv T^{(n_{_{\rm
max.}})}$), and alternative field equations \ngeea\ as
$$
\Re_{\alpha\sbeta}={1\over 2}\kappa^2\, \cs_{\alpha\sbeta}\ ,\eqn\afge$$
where $\cs_{\alpha\sbeta}\equiv S^{(n_{_{\rm max.}})}_{\alpha\sbeta}$.
We will refer to $\cs_{\alpha\sbeta}$ and $\ct$ as generalized Source tensor
and trace generalization of the energy-momentum tensor respectively.
\par
Finally, for a better view into the separation of $\Re$ into $T$ and $T'$,
one can expand the summations in their relevant equations as follows:
$$
\eqalign{\Re
  &\equiv \sum_{0<n<{D\over 2}}\, c_n\, R^{(n)}=c_1\, R^{(1)}+c_2\, R^{(2)}
   +c_3\, R^{(3)}+\cdots\cr
  &\equiv {-\kappa^2\, T\over D-2}+{-\kappa^2\, T'\over D-2}
   \ \qquad{}\qquad\quad {\rm by\ Eq.}\ \dgt\cr
  &=\!\!\sum_{0<n<{D\over 2}}\!{D-2n\over n\bigl(D-2\bigr)} c_n R^{(n)}\!
   +\!\!\sum_{0<n<{D\over 2}}\!{D(n-1)\over n\bigl(D-2\bigr)} c_n R^{(n)}
   \quad {\rm by\ Eqs.}\, \TraceG\, \&\, \dtn\cr
  &=\biggl[c_1\, R^{(1)}+{D-4\over 2\bigl(D-2\bigr)}\, c_2\, R^{(2)}
   +{D-6\over 3\bigl(D-2\bigr)}\, c_3\, R^{(3)}+\cdots\biggr]\cr
  &\quad +\biggl[{D\over 2\bigl(D-2\bigr)}\, c_2\, R^{(2)}
   +{2\, D\over 3\bigl(D-2\bigr)}\, c_3\, R^{(3)}+\cdots\biggr]\ ,\cr}
   \eqn\tandtp$$
where $n_{_{\rm max.}}$ is related to $D$ by equation \nlim . As can be
seen, $R^{(1)}$ does not contribute to $T'$, and higher order gravities
contribute more to $T'$ the higher their orders, i.e.
$\bigl(0,\, {1\over 2},\, {2\over 3},\, \ldots ,\, {n-1\over n}\bigr)D$,
see equation \dtn .
\par
For example, if $D=8$, then $n_{_{\rm max.}}\!=3$\ and we have
$$
\eqalign{\Re
  &=c_1\, R^{(1)}+c_2\, R^{(2)}+c_3\, R^{(3)}\cr
  &\equiv -{1\over 6}\, \kappa^2\, T-{1\over 6}\, \kappa^2\, T'\cr
  &\equiv \biggl(c_1\, R^{(1)}+{1\over 3}\, c_2\, R^{(2)}
   +{1\over 9}\, c_3\, R^{(3)}\biggr)\cr
  &\quad +\biggl(\,{2\over 3}\, c_2\, R^{(2)}+{8\over 9}\,
   c_3\, R^{(3)}\biggr)\ .\cr}$$
\section{\bf Discussions}
\indent
The following table summarizes our results of the analogous demands between
Einstein's and Lovelock's gravitational theory.
\nextline \centerline{\tenpoint {\bf Table 1}:\ Analogy between
                                                Einstein's and
                                                Lovelock's gravitational theory.}
{\tenpoint
$$
\vbox{\offinterlineskip
  \hrule
  \halign{&\vrule#&\strut\quad\hfil#\hfil\quad
          &\vrule#&\strut\quad\hfil#\hfil\quad
          &\vrule#&\strut\quad\hfil#\hfil\quad&\vrule#\cr
    height2pt&\omit&&\omit&&\omit&\cr
    &&&{\bf Einstein's Theory}&&{\bf Lovelock's Theory}&\cr
    height2pt&\omit&&\omit&&\omit&\cr
    \noalign{\hrule}
    height2pt&\omit&&\omit&&\omit&\cr
    &Field equations:&&$G_{\alpha\ssbeta}={1 \over 2}\kappa^2\,
                      T_{\alpha\ssbeta}$&&$\scg_{\alpha\ssbeta}={1 \over 2}
                                          \kappa^2\, T_{\alpha\ssbeta}$&\cr
    height2pt&\omit&&\omit&&\omit&\cr
    &where:&&$G_{\alpha\ssbeta}\equiv R_{\alpha\ssbeta}-{1 \over 2}\,
            \sbg_{\alpha\ssbeta}\, R$&&$\scg_{\alpha\ssbeta}\equiv
                                       \Re_{\alpha\ssbeta}-{1 \over 2}\,
                                       \sbg_{\alpha\ssbeta}\,\Re$&\cr
    height2pt&\omit&&\omit&&\omit&\cr
    \noalign{\hrule}
    height2pt&\omit&&\omit&&\omit&\cr
    &Alternative forms:&&$R_{\alpha\ssbeta}={1\over 2}\kappa^2\,
                       S_{\alpha\ssbeta}$&&$\Re_{\alpha\ssbeta}={1\over 2}
                                           \kappa^2\,\scs_{\alpha\ssbeta}$
                                           &\cr
    height2pt&\omit&&\omit&&\omit&\cr
    &where:&&$S_{\alpha\ssbeta}\equiv T_{\alpha\ssbeta}-{1 \over D-2}\,
            \sbg_{\alpha\ssbeta}\, T$&&$\scs_{\alpha\ssbeta}\equiv
                                       T_{\alpha\ssbeta}-{1 \over D-2}\,
                                       \sbg_{\alpha\ssbeta}\,\sct$&\cr
    height2pt&\omit&&\omit&&\omit&\cr
    &and:&&$T=-\kappa^{-2}\,(D-2)R$&&$\sct\equiv -\kappa^{-2}\,(D-2)\Re$&\cr
    height2pt&\omit&&\omit&&\omit&\cr
    &but:&&. . .&&$\sct=T+T'$&\cr
    height2pt&\omit&&\omit&&\omit&\cr
    &where:&&. . .&&$T=-\kappa^{-2}\,\sum\bigl({D\over n}-2\bigr)c_n\,
                   R^{(n)}$&\cr
    height2pt&\omit&&\omit&&\omit&\cr
    &&&&&$T'\equiv -\kappa^{-2}\,D\,\sum{n-1\over n}\, c_n\, R^{(n)}$&\cr
    height2pt&\omit&&\omit&&\omit&\cr}
  \hrule}$$ }
\indent
As it is obvious from Table~1, there still remains some
(slight) differences in our developing structure (which is based
on analogy).
\par
The effect of $T'$ is what that is zero in Einstein's gravitational
theory and is included only in higher order gravitational theories. This
has also been verified in the quantum theory, as
{\it trace anomalies}\rlap.$^{^{[\dufc]}}$\
In the quantum theory, one relates and justifies the presence of the trace
anomaly, classically, to higher order gravities\rlap.$^{^{[\dufb]}}$\ Here,
we have actually stated a classical view of gravitation which
explicitly shows the presence of an extra (anomalous) trace for the
energy-momentum tensor with an indication of the constitution of higher
order gravities towards this trace anomaly (see equation~\tandtp ).
In the next section, as an example, we will use this procedure for any
generic second order Lagrangian to derive trace anomaly relation suggested
by Duff\rlap.$^{^{[\dufa]}}$
\par
In addition, one may speculate, using the analogous demand, that
in those cases where higher order gravities dominate, space-time
``behaves'' as if its energy-momentum has been ``transferred''
into matter's energy-momentum in the sense that in a universe
devoid of ``matter'' there should be also no meaning for the
existence of space-time, i.e. a strong version of Mach's
principle. Hence, if one adopts the view that the geometrical
curvature induces matter\rlap,\foot{Also note that, in Einstein's
gravity, true gravitational fields exist even in empty space, i.e.
their source should be the space-time itself.}\ then, when
$T_{\alpha\sbeta}$ vanishes, by equation \afge , the gravitational
field equations are
$$
\Re_{\alpha\sbeta}={1\over 2}\kappa^2\Bigl(-{1\over D-2}\Bigr)
                               \bg_{\alpha\sbeta}\, T'\ ,\eqn\empge$$
and thereby seen to be more consistent with Machian
ideas\rlap.\foot{Note that, we are considering the {\it complete} Lovelock's
gravity (as a {\it generalized} Einstein's gravity) that is here compared
with Einstein's gravity. Whereas, for example, even $\Re_{\alpha\ssbeta}=0$
is obviously different from $R_{\alpha\ssbeta}=0$, as the former is
$R_{\alpha\ssbeta}+R^{(2)}_{\alpha\ssbeta}+\cdots=0$.}\
Usual matter just modifies an already existing space-time but the
``matter'', which we described above, would be the only source for the
structure of space-time.
\par
It seems that the above view violates the fundamentals of establishing a
gravitational theory in the sense that in its field equation,
$$
({\rm gravitation\ tensor})\propto ({\rm source\ term})\ ,$$
the left hand side should represent the functions of geometry and the right
hand side should denote source terms independent of the gravitational
characters. However, the ``real'' or ``usual'' physical phenomena, to which
one refers, corresponds to an energy scale of less than the Planck energy,
i.e.
the low energy approximation scale in the superstring theory. Also, the
applicability of higher order gravitational theories are restricted only by
the energy scale. Therefore, generally speaking, these theories do not have
to satisfy all the requirements imposed on the fundamental theory. In other
words, as the coefficients of higher order gravities are very small, one
cannot detect such their implications in ``real'' world. However, these
effects are important in highly curved areas, such as the very early
universe, or in quantum physics.
\par
On the other hand, in the literature, there are also the following views:
\par
\frenchspacing
\item{(a)} The field equations for (apparently) empty higher dimensions are,
           in fact, equations for gravity with matter in lower
           dimensions\rlap.$^{^{[\sal]}}$
\item{(b)} A large class of higher order theories of gravitation are
           equivalent to general relativity plus additional matter fields
           with a new metric (often referred to as ``dynamical universality''
           of Einstein's gravity, see, for example, Ref.~[\mffabsirz] and
           references therein).
\item{(c)} There are conjectures about the origin of inertia in which
           inertia, as in the Machian idea, is not an intrinsic property of
           matter (see, for example, Ref.~[\hrpmashabb]).
\section{\bf Derivation of Duff's Suggested Relation}
\indent
The covariant property of equations is our main theme of search. That is,
by analogy, we demand that the form of the linear Lagrangian theory of
Einstein is left unchanged when one works with the non-linear Lagrangian
theory of gravitation, i.e. the form of equations \elove\ and \grt\
must holds for any such a Lagrangian term\rlap.$^{^{[\farb]}}$
In this section, we take this analogy to any generic second order
Lagrangian, i.e. any combination of geometrical scalar Lagrangian terms
with the HDN of plus two.
\par
The only possible such terms are
$$
R^2 ,\quad\! R_{\mu\nu}\, R^{\mu\nu} ,\quad\! R_{\mu\nu\rho\tau}\,
R^{\mu\nu\rho\tau} ,\quad\! R_{;\,\mu}{}^{\mu}\ ,\quad\!
R^{\mu\nu}{}_{;\,\mu\nu}\ ,\quad\! R_{\mu\nu\rho\tau}\,
{}^\ast\!R^{\mu\nu\rho\tau} ,\quad\! R_{\mu\nu\rho\tau}\,
{}^{\ast\ast}\!R^{\mu\nu\rho\tau} ,$$ where
${}^\ast\!R^{\mu\nu\rho\tau}$ is the dual of $R^{\mu\nu\rho\tau}$
which in four dimensions is defined as
$$
{}^\ast\!R^{\mu\nu\rho\tau}\equiv {1\over 2}\, R^{\mu\nu}{}_{\alpha\sbeta}\,
  {\widetilde \varepsilon}^{\alpha\sbeta\rho\tau}\ ,$$
and where ${\widetilde \varepsilon}^{\alpha\sbeta\rho\tau}$ is the
Levi--Civita tensor density.
\par
But, the fourth and fifth terms give no contribution to the
variation of the action, as they are complete divergences. The
Lagrangian $R_{\mu\nu\rho\tau}\, {}^\ast\!R^{\mu\nu\rho\tau}$, due
to the Bianchi identity, reduces to a boundary term and vanishes
identically\rlap.$^{^{[\lanb]}}$ Also, the last term is actually
equal to$^{^{[\lanb]}}$
$$
R_{\mu\nu\gamma\delta}\, {}^{\ast\ast}\!R^{\mu\nu\gamma\delta}=
  -\Bigl(R^2-4 R_{\mu\nu}R^{\mu\nu}+R_{\mu\nu\rho\tau}\,
  R^{\mu\nu\rho\tau}\Bigr)\ ,$$
where this is the second term of the Lovelock Lagrangian,
the well known Gauss--Bonnet combination or the so-called Lanczos
Lagrangian\rlap,$^{^{[\lanb]}}$\ which in {\it four} dimensional space-time
is a topological invariant$^{^{[\kono]}}$\ and
its variation with respect to the metric leads only to a complete
divergence, and hence, vanishes identically (the so-called
Gauss--Bonnet theorem).
\par
Hence, the corresponding generic Lagrangian, in $D\geq 3$
dimension\rlap,\foot{Obviously, in three and four dimensions, only
two of these three terms are effective.} is
$$
L^{(2)}_{\rm generic}={1 \over\kappa^2}\Bigl(a_1 R^2+a_2 R_{\mu\nu}
       R^{\mu\nu}+a_3 R_{\alpha\hbox{$\beta \textfont1=\seveni$}\mu\nu}\,
       R^{\alpha\hbox{$\beta \textfont1=\seveni$}\mu\nu}\Bigr)\ ,\eqn\lii$$
and its corresponding Euler--Lagrange expression is
$$
\eqalign{G^{(2)}_{({\rm generic})\alpha\hbox{$\beta \textfont1=\seveni$}}
           \!&\equiv {\kappa^2\over \sqrt{-g}}\,{\bdelta\bigl(L^{(2)}_{\rm
              generic}\sqrt{-g}\bigr)\over \bdelta\bg^{\alpha\sbeta}}\crr
           =&2\biggl[a_1 R\, R_{\alpha\hbox{$\beta \textfont1=\seveni$}}
               -2a_3 R_{\alpha\mu}\, R_{\hbox{$\beta \textfont1=\seveni$}}
               {}^{\mu}+a_3 R_{\alpha\rho\mu\nu}\, R_{\hbox{$\beta
               \textfont1=\seveni$}}{}^{\rho\mu\nu}
               +\bigl(a_2+2a_3\bigr)R_{\alpha\mu\sbeta\nu}\, R^{\mu\nu}\cr
              &\quad -\bigl(a_1+{1\over 2}a_2+a_3\bigr)\,
               R_{;\,\alpha\hbox{$\beta \textfont1=\seveni$}}
               +\bigl({1\over 2}a_2+2a_3\bigr)\,
               R_{\alpha\hbox{$\beta \textfont1=\seveni$};\,\mu}{}^{\mu}
               \biggr]\cr
              &\!-{1\over 2}\, \hbox{\fourteeni g}_{\alpha\hbox{$\beta
               \textfont1=\seveni$}}\biggl[\Bigl(a_1 R^2+a_2 R_{\mu\nu}
               R^{\mu\nu}+a_3 R_{\lambda\rho\mu\nu}\,
               R^{\lambda\rho\mu\nu}\Bigr)
               -\!\bigl(4a_1+a_2\bigr)R_{;\,\mu}{}^{\mu}\biggr],\cr}
               \eqn\nlii$$
where it is obviously of the form of equation \elove , i.e.
$$
G^{(2)}_{({\rm generic})\alpha\sbeta}\equiv
R^{(2)}_{({\rm generic})\alpha\sbeta}
-{1\over 2}\bg_{\alpha\sbeta}\, R^{(2)}_{\rm generic}\ .\eqn\gtgeneric$$
\par
Now, the form of relation \grt , i.e. the ``trace'' relation
$$
{\rm Trace}\, R^{(2)}_{({\rm generic})\alpha\hbox{$\beta
            \textfont1=\seveni$}}=
            R^{(2)}_{\rm generic}\ ,\eqn\rTii$$
holds trivially, for non-zero coefficients, if and only if
$$\Boxed{
3\, a_1+a_2+a_3=0}\ .\eqn\cnlii$$
\par
Therefore, the analogy of Einstein's gravitational theory, including the
``trace'' relation, cannot be achieved for {\it any} combination in the
generic Lagrangian case of $L^{(2)}_{\rm generic}$, unless
the above relation between its constituent coefficients
are valid. That is, when its constants are not all independent, but obey
the above constraint, which gives two degrees of freedom, in $D > 4$,
to choose constituent coefficients.
\par
The Lanczos Lagrangian does indeed satisfy the above condition, as expected.
So, in this sense, one may say that the coefficients used in the
Lanczos Lagrangian are specific to a equivalence class of any generic
second order Lagrangians
having the same analogy including the ``trace'' relation. However, the
field equations of all these Lagrangians are fourth order with respect to
the metric, except the Lanczos Lagrangian which is of second order.
\par
Note that, when one considers a {\it homogeneous} Lagrangian, e.g.
$L^{(2)}_{\rm generic}$, {\it alone}, then obviously, one gets its
corresponding {\it homogeneous} Euler--Lagrange terms with a {\it uniform}
HDN. And therefore, one can work with the usual {\it trace} instead of the
{\it Trace} operator, and obtains the same results if one demands the
appropriate ``trace'' relation, i.e.
${1\over n}\, {\rm trace}\, R^{(n)}_{({\rm generic})\alpha\sbeta}
 =R^{(n)}_{\rm generic}$. Though, the notion of Trace operator was
introduced as to be able to deal with when one considers the
Einstein--Hilbert Lagrangian plus higher order terms, i.e. when one works
with an inhomogeneous Lagrangian (see Ref.~[\farb] for details).
\par
As mentioned before, the constraint \cnlii\ is related to the relevant
constraint in the trace anomaly of the energy-momentum tensor. We first
give a brief review of this subject and then explain the above relation.
\par
In the absence of a viable theory of quantum gravity, quantum field theory
in a curved background space-time has been used as the starting point for an
approach to a complete quantization of gravity\rlap,$^{^{[\bida]}}$\ though
one may still argue about an internal inconsistency\rlap.$^{^{[\dufaa]}}$\
In this procedural attempt for quantum aspects of gravity,
while the Einstein gravitational field is retained as an unquantized,
classical background curved metric, the matter fields are treated quantum
mechanically and quantized in the usual way.
\par
In this approach, one sets
$\bg_{\mu\nu}=\bg^c_{\mu\nu}+{\overline {\bg}}_{\mu\nu}$,
where $\bg^c_{\mu\nu}$ is the classical metric of Einstein's gravitational
background space-time, and ${\overline {\bg}}_{\mu\nu}$ is taken to be a
quantum field propagating in this background. Then, in this semi-classical
theory, {\it regularization} techniques are used to compute a finite,
renormalized quantum vacuum expectation value of the energy-momentum tensor,
$\big\langle T_{\mu\nu}\big\rangle=
 {\langle {\rm out},\, 0\mid\, T_{\mu\nu}\mid 0,\, {\rm in}\rangle\over
  \langle {\rm out},\, 0\mid 0,\, {\rm in}\rangle}$.
Where the computation of $\big\langle T_{\mu\nu}\big\rangle$ can also be
treated via its definition by
an {\it effective Lagrangian density}$^{^{[\bos]}}$\
for the quantum matter fields, i.e.
$$
\bdelta{\widetilde L}_{\rm eff}\equiv -{1\over 2}\sqrt{-g}\,
        \big\langle T_{\mu\nu}\big\rangle\, \bdelta\bg^{\mu\nu}
        \ .\eqn\effaction$$
\par
Using the path-integral quantization procedure, one finds$^{^{[\bida]}}$\
that there are divergent terms at the {\it one-loop} level
in the effective action that are local and independent of the state. Hence,
the divergences in the effective Lagrangian, $L_{\rm div}$, are
entirely geometrical, and are built out of local tensors, i.e., actually,
the higher order Lagrangians. These are interpreted as a contribution to
the gravitational rather than the quantum matter Lagrangian. Although, as
the coefficients of these terms diverge as $1/(D-2n)$ in the
{\it dimensional continuation} method of regularization, in odd dimensions
$L_{\rm eff}$ is finite, and hence there is {\it no} anomaly in odd
dimensional space-times. So, one must therefore introduce, {\it a priori},
the relevant counterterms into the original Einstein's Lagrangian with
{\it bare} coefficients into which the divergent terms can be absorbed to
yield renormalized coefficients\rlap.$^{^{[\stela]}}$\ Hence, the
{\it renormalized effective Lagrangian},
$L_{\rm ren}\equiv L_{\rm eff}-L_{\rm div}$, are then finite.
\par
The inclusion of higher order Lagrangian terms have also appeared
in the effect of string theory on classical gravitational physics
by means of a low energy effective action which expresses gravity
at the classical level\rlap.$^{^{[\chswgswltp]}}$\ This effective
action in general gives rise to fourth order field equations (and
brings in ghosts), and in particular cases, i.e. in the form of
dimensionally continued Gauss--Bonnet densities, it is exactly, as
mentioned at the beginning, the Lovelock terms (and consequently
no ghosts arise)\rlap.$^{^{[\zwizum]}}$\ The Gauss--Bonnet term
also appears naturally in the next-to-leading order term of the
heterotic string effective action and plays an essential role in
the Chern--Simons gravitational theories\rlap.$^{^{[\chammuh]}}$
\par
Now, in the special cases when the classical action of the matter field
is invariant under conformal transformations, e.g. in the conformally
coupled massless scalar field, the effective action is$^{^{[\bida]}}$\ also
conformally invariant. But, it has been shown$^{^{[\bida]}}$\ that before
relaxing the regularization, the divergent term in the effective action,
away from $D=2n$ dimension, is {\it not} conformally invariant, although it
is in the physical limit $D=2n$. Apparently, an indication of this conformal
breakdown survives in the physical quantities even when one relaxes the
regularization at the end of the calculation.
\par
In any case, the anomalies generally occur in any regularization method as
a consequence of introducing a scale into the theory in order to regularize
it\rlap.$^{^{[\bida]}}$\ The contribution of $L_{\rm div}$ to the trace of
the energy-momentum tensor,
$\big\langle T_\rho{}^\rho\big\rangle_{\rm div}$, is one of the above
consequences. Hence, the finite, renormalized
$\big\langle T_{\mu\nu}\big\rangle_{\rm ren}$ must also have a nonvanishing
trace, i.e.
$\big\langle T_\rho{}^\rho\big\rangle_{\rm ren}=-\big\langle
         T_\rho{}^\rho\big\rangle_{\rm div}$,
despite the fact that the classical energy-momentum tensors, for the
conformally invariant classical actions, must be traceless.
This is known as a {\it conformal}, or {\it trace}, or {\it Weyl},
anomaly. Note that, as ${\Roman 1}_{\rm eff}$ itself is a conformally
invariant action, the expectation value of the trace of the {\it total}
energy-momentum tensor is zero.
\par
It is now known$^{^{[\bida]}}$\ that all
the regularization techniques give conformal anomalies for fields of
arbitrary spin, but with the same value for the scalar field, and even in
non-conformal theories the anomalies still survive, and in addition are
accompanied by contributions through the lack of conformal invariance.
It has also been clarified$^{^{[\dufc]}}$\ that the anomalous
terms cannot be removed and hence the conformal anomalies are inevitable.
Besides, the anomalies which cannot be absorbed by a finite local
renormalization have also been referred to as {\it non-local} conformal
anomalies\rlap.$^{^{[\ddi]}}$\
\par
Therefore, the Weyl conformal invariance displayed
by classical (massless) fields in interaction with Einstein's gravity no
longer survives in the quantum theory, and is generally broken as a
consequence of the appearance of the Weyl anomalies at the one-loop level.
So, the Weyl conformal invariance, perhaps, is not a {\it good} symmetry
beyond the classical level.
\par
It has been shown$^{^{[\ddi , \chr]}}$\ that, the most general
form of the anomalous trace of the energy-momentum tensor for
classically conformally invariant fields of arbitrary spin and
dimension, when terms only quadratic in the Riemann-Christoffel
tensor and its contractions are considered, is
$$
\big\langle T_\rho{}^\rho\big\rangle_{\rm ren}=-{\hbar\, c\over 180(4\pi)^2}
       \Bigl(a_1 R^2+a_2 R_{\mu\nu}
       R^{\mu\nu}+a_3 R_{\alpha\sbeta\mu\nu}\, R^{\alpha\sbeta\mu\nu}
       +\delta\, R_{;\,\rho}{}^\rho\Bigr)\ ,\eqn\fanom$$
where numerical coefficients $a_1$, $a_2$, $a_3$ and $\delta$ are
dimensionless constants which
also determine the counterterms and are expected$^{^{[\duf]}}$\ to be unique.
\par
In the process of re-examining the Weyl anomaly's applications,
Duff noticed$^{^{[\dufa]}}$\ that, there is a consistency
condition on these numerical coefficients when the dimensional
regularization is applied to a classically conformally invariant
theory in arbitrary dimension. That is
$$\Boxed{
4\, a_1+a_2=a_1-a_3=-\delta}\ ,\eqn\dufeq$$ which holds for
several explicit calculations in the case of conformal scalars,
massless spin-${1\over 2}$ fermions and spin-one gauge fields in
an external gravitational field. The calculations by most authors
confirmed$^{^{[\dufa]}}$\ the above constraint for the values of
$a_1$, $a_2$ and $a_3$ in all cases, but not always that of
$\delta$. Apparently, this is due to the fact that the term
$R_{;\,\rho}{}^\rho$ is a local anomaly\rlap.$^{^{[\dufa]}}$\
However, the absorption of this term by any of the above mentioned
four dimensional actions breaks the conformal invariance of the
action, and hence, the above constraint cannot be applied.
\par
The first part of the above consistency condition \dufeq\ was later
derived$^{^{[\bcr , \bpb]}}$ based on a {\it cohomological} point of
view, using the Wess--Zumino consistency conditions, which was
claimed to be the true reason for the existence of such a relation.
And more recently, there are some works in the literature which
claim that the AdS/CFT correspondence, namely the holographic
conformal anomaly, maybe responsible for it, see e.g.
Ref.~[\noodog]\ and references therein.
\par
In the previous sections, we argued that, based on the analogous
demand, there is a trace generalization of the energy-momentum
tensor i.e., $\ct =T+T'$, and actually, we stated a classical view
of gravitation which explicitly shows the presence of an extra
trace for the energy-momentum tensor, namely $T'$ equation~\dtn .
The same argument for generic cases yields
$$
T'=-\kappa^{-2}\,D\,\sum_{n\geq 1}\,{n-1\over n}\, c_n\,
   R^{(n)}_{\rm generic}\equiv \sum_{n\geq 1}\, T'_n\ ,\eqn\dtng$$
where $R^{(1)}_{\rm generic}\equiv R$\ (however, $T'_1=0$), and there is no
upper limit for $n$.
\par
In this section, we considered the case of $n=2$, and in this case
one only has
$$
T'_2=-{\kappa^{-2}\,D\over 2}\, c_2\, R^{(2)}_{\rm generic}\
,\eqn\dttg$$ where $R^{(2)}_{\rm generic}$, according to
equations~\nlii\ and \gtgeneric , is
$$
R^{(2)}_{\rm generic}\equiv \kappa^2\,L^{(2)}_{\rm generic}-\bigl(4\, a_1
                               +a_2\bigr)R_{;\,\mu}{}^{\mu}\ ,\eqn\rgii$$
with constraint~\cnlii . Comparing it with equation~\fanom , it
obviously shows that
$$
\delta=-\bigl(4\, a_1+a_2\bigr)\ .\eqn\dodelta$$
Hence, it completely gives the trace anomaly relations suggested by Duff.
Also, by matching equation~\fanom\ with equation~\dttg\ in $D=4$ dimensions,
one gets $c_2={1\over 90(4\pi)}\ell_p{}^2$, as expected.
\par
In the above mentioned semi-classical theory, the effective action
is$^{^{[\avra]}}$\ a {\it covariant functional} i.e., invariant under
diffeomorphisms and local gauge transformations. Therefore, the
approximation procedures for calculating the effective action have to
preserve the general covariance at {\it each order}. Hence, conformal
invariance is also sacrificed to the needs of general covariance. This is
what we have actually performed in the classical theory of gravitation
through preserving the covariant property of the linear Lagrangian theory
of Einstein's gravity for each order of non-linear Lagrangian theories of
gravitation, by an analogous demand.
\par
Hence, the origin of Duff's suggested relation, equation~\dufeq ,
between the coefficients of the conformal anomalies may classically
be interpreted due to the general covariance of Einstein's theory.
Though, it is somehow a naive conjecture, nevertheless, it gives
almost an easy classical manner to grasp its result.
\par
Taking the analogy, that is the existence of the trace anomaly of
the energy-momentum tensor even in classical treatments, further to
any type of the third order Lagrangian$^{^{[\farf]}}$\ gives a more
deep sight of this procedure. Actually, investigation
shows$^{^{[\fard]}}$\ that there is an interesting similarity
between Weyl invariant combinations in six dimensions$^{^{[\bpb ,
\desch]}}$\ and the constraint relation that coefficients of any
generic third order Lagrangian must satisfy in order to hold the
desired analogy, which then, can be used as a criterion. This may
also lead one to speculate that there should be an intrinsic
property between the appropriate heat kernel
coefficients$^{^{[\gilababocsch]}}$\ and the covariance of the form
of the Einstein equations, and or between the matter and the
geometry. Probing these thoughts, by further extending the analogous
procedure, to see whether there is a geometrical meaning behind this
generalization, perhaps besides what have already been given in the
literature\rlap,$^{^{[\bcr , \bpb ,\noodog]}}$\ has been our most
important task on which our search still continues.
\par
\vskip 0.2cm
{\bf Acknowledgment}:\ The author is grateful to Prof. John M Charap and
                       the Physics Department of Queen Mary \& Westfield
                       College University of London where some part of this
                       work has been carried out.
\tenpoint
\refout
\bye